\documentstyle[12pt]{article}
\begin{document}
\title{Self-consistency and Symmetry \\ in \\ $d$-dimensions}
\renewcommand{\thefootnote}{\fnsymbol{footnote}}
\author{Serge Galam  \footnotemark[1]\\  (E-mail: galam@gps.jussieu.fr)\\ $\,$\\
Acoustique et Optique de la Mati\`{e}re Condens\'{e}e \footnotemark[2]\\
Tour 13 - Case 86, 4 place Jussieu, \\ 75252 Paris Cedex 05, France\\}
\date{To appear in Phys. Rev. B (1996)}
\maketitle

\addtocounter{footnote}{1}
\footnotetext{Permanent Adress: Groupe de Physique des Solides, Universit\'e Paris 7, Tour 23,\\
\hspace*{0.6cm}2 place
Jussieu,
75251 Paris Cedex 05.
Laboratoire associ\'{e} au CNRS (URA n$^{\circ}$ 17)}
\addtocounter{footnote}{1}
\footnotetext{Laboratoire associ\'{e} au CNRS (URA n$^{\circ}$ 800)}

\begin{abstract}

Bethe approximation is shown to violate Bravais lattices translational invariance.
A new scheme is then presented which goes over the one-site Weiss model yet
preserving initial lattice symmetry.
A mapping to a one-dimensional finite closed chain in an external field is obtained.
Lattice topology determines the chain size. 
Using recent results in percolation, lattice connectivity between chains is argued to be 
$\frac{q(d-1)-2}{d}$
where $q$ is the coordination number and $d$ is the space dimension. A new 
self-consistent mean-field equation of state is derived. Critical 
temperatures are thus calculated for a large variety of lattices and dimensions. Results are
within a few percent of exact estimates.
Moreover onset of phase transitions is found to occur in the range $(d-1)q> 2$.
For the Ising hypercube it yields the Golden number limit $d > \frac{1+\sqrt 5}{2}$.

\end{abstract}
\newpage

\section{Introduction}
The Weiss theory of ferromagnetism, also known as the mean-field theory, has
been proposed almost a century ago [1]. Its major achievement is to provide a
simple and universal model for collective phenomena and phase transitions.
In most cases it yields phase diagrams which are qualitatively correct.
However, quantitative results are very poor. In particular, all aspects of
the critical behavior are grossly misrepresented.

Thirty years latter, Bethe extended the Weiss one-spin treatment to a
cluster of $(q+1)$ fluctuating spins where $q$ is the coordination number [2]. For the one-dimensional
Ising model, this scheme reproduces the exact results. At higher dimensions,
though some non-universal critical parameters are better represented, the
critical behavior stays classical. 

However it was recently pointed that within a Bethe scheme different site magnetizations 
are produced respectively in the $(q+1)$ cluster and its surrounding [3]. Therefore
the Bethe scheme violates Bravais lattices translational invariance.
This symmetry violation which has been overlooked for decades, is indeed consistent with
a well known result. Bethe
equations of state correspond to the exact solution of a model on the deep
interior of a Cayley tree (which by definition, has no translational symmetry and exhibits singularities
at its surface [4]). 

In this paper we  present a new self-consistent approach which embodies 
Bethe ideas, including fluctuations of a spin cluster, but
preserves the lattice translational symmetry.

The outline of this paper is as follows. A generalized lattice version of
the usual mean field treatment of the Ising ferromagnet is presented in
Section 2. The underlying use of symmetry is thus explicited.
Along the same lines, the Bethe scheme is revisited in
Section 3. It shows the initial lattice symmetry is always broken.
Section 4 is 
devoted to the presentation of a new self-consistent treatment which does preserves the 
symmetry. The lattice is divided into finite closed loops of fluctuating spins and 
mean-field spins respectively. The problem is thus mapped onto
a one-dimensional closed fluctuating chain in an external field which allows
an exact solution. The field is
$h\equiv \delta Jm$ where  $J$ is the 
coupling exchange, $m$ is the mean-field average spin magnetisation and $\delta$ is 
the chain connectivity. 
Using very recent
results on percolation [5]
it is argued that $\delta =\frac{q(d-1)-2}{d}$
where $d$ is the space dimension. Critical temperatures are
calculated in Section 5 for a large variety of
lattices at several dimensions. Discrepancies with available exact estimates are within 
few percent. In the last Section
phase transitions are found to occur only in the range $(d-1) q>2$ for $d$-dimensional 
Bravais lattices with coordination $q$. In the
hypercubic ferromagnetic Ising case it gives the Golden number limit 
$d > \frac{1+\sqrt 5}{2}$.

\section{The mean field scheme}

Mean-field theory is basically a one-site approach. However we present here a new
generalized lattice scheme which enhances its interplay with symmetry. 
We illustrate it in the case of a ferromagnetic Ising system on a lattice with
q nearest neighbors. The  Hamiltonian is,
\begin{equation}
H=-J\sum_{(i,j)}S_{i}S_{j}\:.
\end{equation}
where $S_{i}=\pm1$ is an Ising variable, and interactions $J$	are restricted 
to nearest neighbors (nn). 
A mean-field treatment of Eq. (1) processes indeed in two steps. 
\subsection{Breaking the lattice symmetry}
 Two interpenetrated lattices $A$ and $B$ are
created within a staggered symmetry. Thermal fluctuations are then ignored on 
the B-sublattice. Spins $S_{j}^{B}$ are substituted to their common thermal average 
$<S_{j}^{B}>$ denoted $m\equiv<S_{j}^{B}>$.
At this stage spins $S_i^{A}$ are no longer equivalent to their $q$ nn $S_{j}^{B}$. 
The translational lattice symmetry has been broken. Eq. (1) turns to,
\begin{equation}
H_{MF}=-J\sum_{i\in A}S_{i}^{A}\sum_{j\in B,j=1}^{q}<S_{j}^{B}>\ ,
\end{equation}
which is a decoupled one-site Hamiltonian. 
The partition function can now be calculated integrating
over $S_i^{A}=\pm1$. Magnetization at site $i\in A$ is then, 
\begin{equation}
<S_{i}^{A}>=\tanh \{K qm\}\ ,
\end{equation}
where $K\equiv \beta J$, $\beta \equiv \frac{1}{k_BT}$, $k_B$ is the Boltzmann constant 
and $T$ the temperature.
\subsection{Restoring the symmetry}
At this stage symmetry has been artificially broken 
through 
the introduction of two interpenetrated sublattices $A$ and $B$. 
The second step in the mean-field scheme is then to restore the initial lattice symmetry. 
It is done 
requiring the equality $<S_{i}^{A}>=m$ which in turn
produces 
the well known self-consistent mean-field equation,
\begin{equation}
m=\tanh\{K qm\}\ .
\end{equation}

Although fluctuations have been included only at one sublattice, the system is again 
homogeneous with 
the same magnetization $m$ at every lattice site. 
The critical temperature $T_c$ is given by,
\begin{equation}
K_c^W=\frac{1}{q}\ .
\end{equation}
Numerical 
estimates from Eq. (5) are rather poor.
For instance, it is $K_c^W=0.25$ for the square lattice ($q=4$) instead of the exact value 
$0.4407$ [6]. 
Moreover $K_c^W=0.50$ at $d=1$ ($q=2$) whereas it should be $T_c=0$.

\section{The Bethe approximation revisited}

We now present an unusual view point about he Bethe scheme to enlighten its symmetry content.
First a finite cluster is considered. Second the Bethe approximation is
implemented starting from a lattice. On this basis it then is shown translational invariance 
is always lost in the process.
\subsection{A cluster in a surface field}

We consider a finite cluster of one spin $S_i$ coupled ferromagnetically to 
$q$ nn spins $S_j$ with an exchange $J$. 
The associated partition function yields $<S_i>= <S_j>=0$ as expected for a finite
spin system. 

An external field $h$ is then applied but only on the surface shell of the cluster,
i. e., only on the $q$ spins $S_j$ and not on the central spin $S_i$. The Hamiltonian is,
\begin{equation}
H_{i}=-JS_{i}\sum_{j=1}^{q}S_{j}-h\sum_{j=1}^{q}S_{j}\:.
\end{equation}  
A non-zero magnetization is induced in the cluster from $h\neq 0$ which is a symmetry
breaking field. 
However since the field is not applied to the center spin, the cluster
is no longer homogeneous, i. e., $<S_i>\neq <S_j>$ with both thermal averages being
a function of $h$.
At order one in $\alpha \equiv \beta h$ we have,
\begin{equation}
 <S_i>=q\tanh(K) \alpha \:,
\end{equation}
and
\begin{equation}
 <S_j>=\{ 1+(q-1)\tanh^2(K)\}\alpha \:.
\end{equation}

We can then ask which value of 
the surface field $h$, if any, makes the 
cluster homogeneous with a non-zero magnetization $<S_i>= <S_j>\neq 0$. 
It leads to the self-consistent 
equation in $\alpha$,
\begin{equation}
\alpha=(q-1)\ln {\frac{\cosh(\alpha +K)}{\cosh(\alpha -K)}}\ .
\end{equation}
Expanding the right-hand side of Eq. (9) around $\alpha =0$ two solutions are obtained. First, 
\begin{equation}
h =k_BT \{3(q-1)(K-K_c)\}^{\frac{1}{2}}\ ,
\end{equation}
for $K \geq K_c^B$ where ($\tanh K_c=\frac{1}{q-1}$),
\begin{equation}
K_c^B= \frac{1}{2} \ln {\frac{q}{q-2}}\ .
\end{equation}
Second we have $h=0$ for $K \leq K_c^B$.

We thus have found that for each value of temperature $K> K_c^B$, there exists a well defined 
non-zero value of $h$ which makes the cluster homogeneous, i. e., $<S_i>=<S_j>\neq 0$. However for 
$K \leq K_c^B$ such a field does 
not exist, i. e., only $h=0$ is compatible with an homogeneous cluster. In this case 
$<S_i>= <S_j>=0$ due to the cluster finiteness. Otherwise $<S_i>\neq <S_j>$.

\subsection{The Bethe implementation}

The Bethe model starts from a lattice in which a $(q+1)$ cluster similar to the one treated above,
is singled out. However the difference from above comes from the definition of the field $h$. 
Above, $h$ is an external magnetic field. Here $h$ is 
argued to represent an effective field which accounts for the coupling between the cluster surface 
and the remainder of the lattice. Calculations are nevertheless exactly identical to those done 
in Subsection (3.1) leading to Eqs. (7) and (8).

It is then usually stated that common magnetization in the cluster, 
i. e., $m\equiv <S_i>= <S_j>$ is the magnetization in the lattice bulk. Eq. (7) gives,
\begin{equation}
m=\frac{q}{q-1}\alpha \ ,
\end{equation}
where $K$ is taken at $K_c$. 

For a good review of the classical presentation of the Bethe model, see the textbook 
of Pathria [7].
Major improvement over  Weiss model is twofold. Values of critical temperatures
are slightly better.
It is $K_c^B=0.35$ instead of $0.25$ for the square case. 
At $d=1$ results are exact with in particular $K_c^B=0$.

\subsection{The Bethe violation of translational invariance}

Nevertheless it has been proven recently [3] that a Bethe implementation on a lattice
breaks translational invariance. To enlight this statement we present here a generalized 
lattice view similar to the one developped above for the mean-field model .

To implement a Bethe scheme on a lattice 
3 distinct interpenetrated sublattices must be introduced artificially.
We start from one fluctuating spin labelled $i$ ($S_{i}$) to which we refer as an A-spin. 
The set of its
nn spins are labelled by index $j$ ($S_{j}^{i}$) and called B-spins. Last, nn spins 
of $j$ spins, except the
$i$ spin, are labelled
by the index $k$ ($S_{k}^{i,j}$) and called  C-spins. 
The A-spin plus these B and C shells constitute a supercell for the lattice. This supercell is 
then used to pave the space and reproduce the whole lattice.
Eq. (1) writes,
\begin{equation}
H=-J\sum_{i\in A}\{S_{i}\sum_{j\in B,j=1}^{q}S_{j}^{i}
+\sum_{j\in B,j=1}^{q}S_{j}^{i}\{\sum_{k\in C}'S_{k}^{i,j}\}\}\:.
\end{equation}
The prime over last sum means the summation over $k$ $j$-nn spins does not includes the $i$-spin. 
The scheme is easily visualized in the square case. For lattices with 
closed loops nn like triangular, Eq. (13) has to be modified to account for interactions 
between nn $S_{j}^{i}$.  

\subsubsection{Breaking the lattice symmetry}

It is worth noticing Eq. (13) is identical to  Eq. (1). At this stage no
approximation was done. Only a formal relabelling has been used. 

Now we perform a mean-field transformation of the Hamiltonian. More remote C-sublattice spins 
$S_{k}^{i,j}$
are  substituted to their thermal average
$m_C \equiv <S_k^{i,j}>$. 
In so doing the lattice symmetry has been artificially broken. However it  
allows a calculation of the partition function.
Eq. (13) reduces to $H=\sum_{i\in A}H_i$ with,
\begin{equation}
H_{i}=-JS_{i}\sum_{j=1}^{q}S_{j}^i-h\sum_{j=1}^{q}S_{j}^i\:,
\end{equation}
where $h\equiv J\sum_{k=1}^{q-1}<S_{k}^{i,j}>\:$ and $k$ runs for all (q-1) nn 
of respectively each $j$ spin, spin $i$ being excluded. Hence
\begin{equation}
h=J(q-1)m_C\:,
\end{equation}
for the effective field. From Eq. (14) we calculate thermal averages $m_A\equiv <S_{i}>$
and $m_B\equiv <S_{j}^{i}>$
as a function of $m_C$.
We end up with three unknown 
magnetizations which are respectively $m_A$, $m_B$ and $m_C$.
\subsubsection{The symmetry breaking is irreversible}

At this stage, as in the mean-field scheme, the artificially broken symmetry has to be restored to
recover the initial physical problem we started from. It is of importance to stress
even a drastic approximation should not change the symmetry of the problem. It means here
the 3 different magnetizations have to be identical.

First the symmetry within the fluctuating cluster A-B can be restored putting $m_A=m_B$. 
This equality 
yields 
a self-consistent equation in $m_C$ identical to Eq. (9),
\begin{equation}
m_C=\frac{1}{K}\ln {\frac{\cosh( K[(q-1)m_C +1])}{\cosh( K[(q-1)m_C -1])}}\ ,
\end{equation}
where $\alpha = K(q-1)m_C$ has been used. 

Next step is to 
extend the cluster homogeneity to the C-sublattice. Denoting $m\equiv <S_i>=<S_j^i>$ 
we have to fulfill the equality $m_C=m$. However Eq. (7) shows that
$m=q\tanh(K)K(q-1)m_C$ 
which in turn obviously demonstrates $m\neq m_C$, as long as Eq. (16) holds.

Alternatively we could have achieved
$m_A=m_C$. Then it is at the expense of getting another self-consitent equation in $m_C$, 
\begin{equation}
m_C=q(q-1)K\tanh(K)m_C \:,
\end{equation}
which is nevertheless not compatible with Eq. (16). In this case $m_A\neq m_B$. 

In both cases, translational invariance is lost.
There exist one super-sublattice A and B (A and C) embedded in the C-sublattice 
(B-sublattice), both having
different magnetizations. We thus have clearly demonstrated
that the Bethe scheme produces always an irreversible lattice symmetry breaking.
\subsubsection{The Bethe scheme is forbidden by symmetry}

Overlooked for several decades, this symmetry breaking  could indeed be 
found readily  from the Bethe scheme cluster character.
Cluster center has all its nn spins which are fluctuating while
surface cluster spins
have $(q-1)$ mean-field nn spins and one fluctuating spin. Simultaneously mean-field spins have 
all their 
nn which are fluctuating spins, making their environnement identical to the cluster center. 
On this basis
the Bethe requiremnt $<S_i>=<S_j>$ is not compatible with the equality $m_C=<S_j>$ which should 
also hold to ensure translational invariance. The Bethe topology is therefore forbidden 
by symmetry [3]. It is not the case for one-site Weiss theory.

Last but not least, it is worth noticing it is 
this very symmetry problem 
which makes the Bethe model exact on the Caley tree lattice.
This lattice does not exhibit translational invariance by construction. However the Cayley tree 
(also called Bethe lattice in the tree's deep interior) is pathological with singularities
at its surface [4].
Spins on the surface have a non-negligible contribution
to the partition function in the thermodynamic limit. This is not the case for a regular lattice.

\section{A preserving symmetry scheme}

Once the Bethe model is found forbidden by symmetry, the underlying question is to find out if a
self-consistent treatment which includes more than one site is indeed possible. We show below
that such a model does exist.
\subsection{A 2-spins model}

Ogushi [8] considered a two nn fluctuating spin cluster.
For one pair $(1,\ 2)$ of fluctuating spins the equality 
$<S_1>=<S_2>$
holds by construction. It allows then to fulfill $<S_1>=m$
as required by lattice
symmetry. Critical temperatures are given by, 
\begin{equation}
(q-1)K^O\{1+\tanh(K^O)\}=1\ .
\end{equation}
Improvement over the Weiss model is only marginal with, for instance,
$K_c^O=0.26$ for the square lattice. Moreover
a transition is still wrongly predicted at $d=1$.
\subsection{A $N$-spin model}

To go over a two-spin cluster while preserving initial lattice symmetry, 
requires to start
with a topology of only two spin species A and B. 
Therefore fluctuating spins $S_i$ make one species A and mean-field spins characterize 
the other species B.
A two-sublattice coverage has to be built up under the constraint of identical A-spins 
and B-spins with respect to their respective topology (nn-surroundings). 
\subsubsection{Breaking the symmetry}
 
Above Bethe 
approximation analysis emphazises the role of the cluster center in the breaking of symmetry. 
Therefore to avoid such a symmetry breaking no fluctuating center should exist within 
the fluctuating cluster. Such a constraint can be
achieved via compact closed linear loops within the lattice topology. For instance
4-spin squares and 3-spin triangles for respectively square and triangular lattices. 

Each one of these plaquettes is then
set respectively as A-species and B-species with
a staggered-like covering pattern. A-plaquettes (B-plaquettes) have thus all their nn plaquettes
as B-plaquettes (A-plaquettes). For a given plaquette, each spin has two nn spins of the same
species within the plaquette itself and $(q-2)$ nn spins of the other species belonging to nn
plaquettes.

At this stage the problem has been turned into to a series of
fluctuating one-dimensional finite closed 
chains in an external field $h$. The number of spins $N$ is determined by the lattice topology.
It is $N=4$, $N=3$, $N=3$ and $N=6$ for respectively square, triangular, 
Kagom$\acute{e}$ and Honeycomb lattices. 
Interactions with nn mean-field spin plaquettes produce the field $h$. We have $h=\delta Jm$ 
where
$\delta$ accounts for connectivety to B-sublattices.

Above transformation allows then an exact solution since the problem of a nn Ising ferromagnet on 
a closed chain in an 
external field can be solved exactly [7]. In particular, the chain site magnetization is,   
\begin{equation}
<S_i>=\beta \exp{2K}\{\frac{(1-\tanh(K)^N)}{(1+\tanh(K)^N)}\}h\ ,
\end{equation}
at order one in $h$. Here $i\in A$-plaquettes.

\subsubsection{Restoring the lattice symmetry}

Then putting $<S_i>=m$ restores initial lattice symmetry. Here symmetry restoring
is possible since  
only two sublattices were involved. It was not the case for the three sublattice Bethe scheme.
The self-consistent equation of state is,
\begin{equation}
m=\delta K \exp{2K}\{\frac{(1-\tanh(K)^N)}{(1+\tanh(K)^N)}\} m+... \ ,
\end{equation}
at order one in $m$ using $h=\delta Jm$.
To solve Eq. (20) we need to determine the value of $\delta$. 
\subsubsection{Rescaling the connectivity}

At this stage it is worth  evoking a recent work on percolation thresholds [5]. It was found that 
relevant connectivity variables for site and bond dilution are repectively
$(d-1)(q-1)$ and $\frac{(d-1)(q-1)}{d}$. In other words, starting from a given site, 
it shows that for site percolation,
the number of possible directions $(q-1)$ has to be multiplied by $(d-1)$.
For bond percolation this effective number of sites has to be divided by dimension $d$.
Using these variables, a universal power law form was found to yield all percolation thresholds
for all lattices at all dimensions [5].

This percolation finding suggests to consider here a rescaled connectivity between closed loops
instead of $\delta =q-2$. Using above counting, we first start with $q$ instead of $(q-1)$
since now dealing with pair exchange interactions and not percolation. Then we
renormalize $q$ by $(d-1)$ giving $q(d-1)$. However the $2$ neighboring
sites which are treated exactly within the closed loop have to be substracted from the effective
 number of sites which gives 
$q(d-1)-2$. Moreover, interactions being related to bonds, we divide this number by $d$ 
as for bond percolation which results in, 
\begin{equation}
\delta =\frac{q(d-1)-2}{d} \ . 
\end{equation}

\section{ Results}
We can now check the validity of our simple symmetry preserving model with respect to critical 
temperatures.
From Eq. (21) we get, 
\begin{equation}
\delta K_c^G \exp{2K_c^G}\{\frac{(1-\tanh(K_c^G)^N)}{(1+\tanh(K_c^G)^N)}\}=1\ .
\end{equation}
The trivial connectivity counting $\delta=q-2$ already improves Weiss model. For instance 
$K_c^G=0.29$ in the square case and $T_c=0$ at $d=1$. We now proceed using Eq. (21) 
for connectivty.
\subsection{$d=2$}

For the square case ($q=4,\ N=4$), $\delta=1$ which gives
$K_c^G=0.4399$. Exact result is $K_c^e=0.4407$. In the case of triangular 
lattice ($q=6,\ N=3$),  $K_c^G=0.2919$ with $\delta=2$ while the exact
estimate is $K_c^e=0.2746$. For Kagom$\acute{e}$ ($q=4,\ N=3$) $\delta=1$ yielding
$K_c^G=0.4649$ for an exact estimate of $K_c^e=0.4666$. And $K_c^G=0.6160$ for the honeycomb 
lattice
($q=3,\ N=6$) where $\delta=\frac{1}{2}$ for an exact estimate of
$K_c^e=0.6585$ (see Table I).

\subsection{$d=3$ }

Going to $d=3$ imposes to restrict the plaquette size to $N=4$ since a 
one-dimensional loop cannot embody a three-dimensional topology. However there exits a $d-$dependence
through Eq. (21). We get $\delta=\frac{10}{3}$, $\delta=\frac{14}{3}$ and 
$\delta=\frac{22}{3}$ for respectively cubic, fcc and bcc lattices.
Corresponding critical temperatures are given by 
$K_c^G= 0.2012,\  0.1568,\  0.1096$ respectively for exact estimates of 
$0.2217,\  0.1575,$ and $0.1021$ (see Table I).

\subsection{$d>3$}

Critical temperature estimates [9, 10] are available for the hypercube at $d=5,\  6,\ 7$. These are
$K_c^e= 0.1139,\ 0.0923,\ 0.0777$ respectively. In comparaison, Eq. (22) gives 
$0.1380,\  0.0869,\  0.0737$. We can also predict values for other lattices like 
for instance $fcc$ ones (see Table I). 

\subsection{$d \rightarrow \infty$}

To get the $d \rightarrow \infty$ asymptotic limit of our model
we take both $q \rightarrow \infty$ and  $J\rightarrow 0$ 
under the constraint $qJ=cst$.
From Eq. (21) connectivity limit is
$\delta \rightarrow q(1-\frac{1}{d})$ which gives always,
\begin{equation}
\delta \rightarrow q \,
\end{equation}
at leading order. Indeed $q$ diverges always quicker than 
$\frac{q}{d}$.even for $fcc$-lattices where $q=2d(d-1)$.  
In turn  Eq. (22) becomes, 
\begin{equation}
K_c^G  =\frac{1}{q}\ ,
\end{equation}
which is exactly Eq. (5). Mean-field 
result is thus recovered as expected in the $d\rightarrow \infty$ limit.

\subsection{$N$-dependence}

To evaluate the sensibility on the loop size, it is fruitful to
expand Eq. (22) in powers of $K$. It gives
\begin{equation}
K_c^G (1+2K_c^G+...+\frac{(2K_c^G)^N}{N!})(1-(K_c^G)^N+...) (1-(K_c^G)^N+...)= \frac{1}{\delta }\ .
\end{equation}
Since $N\geq 3$, a simple analytic expression is obtained only at order one,
\begin{equation}
K_c^G  =\frac{1}{\delta}\ .
\end{equation}
At two dimensions Eq. (26) gives $K_c^G=1,\ \frac{1}{2},\ 1,\ 2$ for respectively the square, 
triangular,  Kagom$\acute{e}$ and for honeycomb lattices.
These results are rather poor and shows the importance of the finite value of $N$ which 
embodies part of the lattice topology.

\subsection{Weiss model with $\delta $ connectivity}

It is interesting to distinguish in our model the linear closed loop contribution from the use
of the rescaled connectivity $\delta $. It is readily done taking back Weiss Equation 
with $\frac{q(d-1)}{d}$ instead of $q$. Here we don't have
to substract the two sites of the linear loop as in Eq. (21).  We thus get,
\begin{equation}
K_c^g=\frac{d}{q(d-1)} \,
\end{equation}
instead of Eq. (5). Critical temperatures at $d=2$ are,
 $K_c^g=0.5000$ for the square ($q=4$, $K_c^{e}=0.4407$), $K_c^g=0.3333$ for triangular 
lattice ($q=6$, $K_c^{e}=0.2746$),
$K_c^g=0.5000$ for Kagom$\acute{e}$ ($q=4$, $K_c^e=0.4666$) and $K_c^g=0.6667$ 
for the honeycomb lattice
($q=3$, $K_c^e=0.6585$) (see Table I).

At $d=3$ we get,
$K_c^g= 0.2500,\  0.1875,\  0.1250$ respectively for exact estimates of 
$0.2217,\  0.1575,$ and $0.1021$ (see Table I).

\section{Conclusion}

We have presented a very simple self-consistent model which both preserves  
initial lattice symmetry and goes beyond the one-site Weiss approach.
It yields values for critical temperatures within a few percent of exact results. 
Besides a rescaled lattice connectivity, the finite
length of the loops is also taken into account. This new scheme represents a substantial
improvement over existing mean-field cluster approximations. However associated critical exponents
stay identical to those of the Weiss model.

Last but not least, we can determine from our model a lower critical dimension for phase 
transitions. 
It comes from the condition $h=0$ for which we have a one-dimensional finite system.
Such a system has  no long 
range order at $T \neq 0$. Phase transitions are thus obtained only in the range $h \neq 0$
which gives, 
\begin{equation}
q(d-1)> 2\ .
\end{equation}
For Ising hypercubes ($q=2d$)
it yields the Golden number limit,
\begin{equation}
d > \frac{1+\sqrt 5}{2}\ ,
\end{equation}
which both excludes the $d=1$ case and contains $d=2$ as it should be.

\subsection*{Acknowledgments.}
The author would like to thank S. R. Salinas for stimulating discussion on the subject.

\newpage
{\LARGE References}\\ \\
1. {\sf P. Weiss}, J. Phys. Radium, Paris \underline {6}, 667 (1907)  \\
2. {\sf H. A. Bethe}, Proc. Roy. Soc. London A\underline {150}, 552 (1935)  \\
3. {\sf S. Galam and A. Mauger}, J. Appl. Phys. \underline {75/10}, 5526
(1994) 

and  Physica A\underline {205}, 502 (1994)   \\
4. {\sf  E. M$\ddot{u}$ller-Hartmann and J. Zittartz}, Phys. Rev. lett. \underline {33}, 893 (1974)  \\
5. {\sf  S. Galam and A. Mauger}, Phys. Rev. B\underline {53}, 2171 (1996)  \\
6. {\sf L. Onsager}, Phys. Rev. \underline {65}, 117 (1944)  \\
7. {\sf R. K. Pathria}, Statistical Mechanics, Pergamon Press (1972)  \\
8. {\sf T. Ogushi}, Prog. Theor. Phys. (Kyoto) \underline {13}, 148 (1955)  \\
9. {\sf M. E. Fisher}, Repts. Prog. Phys. V\underline {XXX} (II), 671 (1967)  \\
10. {\sf  J. Adler}, in ``Recent developments in computer simulation studies 

in Condensed matter physics", VIII, edited by D. P. Landau, 

Springer (1995) \\

\newpage
\begin{table}

\label{tbl}
\begin{center}
\begin{tabular}{|l|l|r|r|r|r|r|}
\hline  
Dimension &Lattice &$\,q$&$\delta$&${K_c^e}$&${K_c^G}$&${K_c^g}$ \\ [5pt] 
\hline 
$d=2$ &Square & 4 &1&  0.4407& 0.4399 &0.5000 \\ [5pt]
$\,$ &Honeycomb & 3&$\frac{1}{2}$&  0.6585& 0.6160&0.6667\\ [5pt]
$\,$ &Triangular & 6&2&  0.2746& 0.2837&0.3333 \\ [5pt]
\hline
\hline
$d=2$ &Kagom$\acute{e}$*&  4&1& 0.4666 & 0.4649&0.5000 \\ [5pt]
\hline
$d=3$ &Diamond & 4& 2&0.3698& 0.2857&0.3750\\ [5pt]
$\,$ &sc& 6&$\frac{10}{3}$&  0.2216& 0.2012&0.2500\\ [5pt]
$\,$ &bcc& 8&$\frac{14}{3}$& 0.1574& 0.1568&0.1875\\ [5pt]
$\,$ &fcc& 12&$\frac{22}{3}$ & 0.1021& 0.1096&0.1250\\ [5pt]
\hline
$d=4$ &sc& 8&$\frac{22}{4}$& 0.1497& 0.1380&0.1667\\ [5pt]
$\,$ &fcc& 24 &$\frac{23}{2}$& $\ $& 0.0749&0.0555\\ [5pt]
\hline
$d=5$ &sc& 10&$\frac{38}{5}$& 0.1139& 0.1064&0.1250\\ [5pt]
$\,$ &fcc& 40&$\frac{158}{5}$ & $\ $& 0.0298&0.0312\\ [5pt]
\hline
$d=6$ &sc& 12&$\frac{29}{3}$& 0.0923& 0.0869&0.1000\\ [5pt]
\hline
$d=7$ &sc& 14&$\frac{82}{7}$& 0.0777& 0.0737&0.0833\\ [5pt]
\hline
\end{tabular}
\end{center}
\caption{\sf $K_c^G$ and $K_c^g$ from this work compared to ``exact estimates" 
$K_c^e$ taken from [9, 10].} 
\end{table}

\end{document}